\documentclass[twocolumn,aps,english,prl,nofootinbib,preprintnumbers]{revtex4}
\usepackage{mathrsfs}
\usepackage{graphicx}
\usepackage{amsmath, amssymb}
\usepackage{babel}
\usepackage{color}
\usepackage{slashed}
\usepackage{hyperref}

\begin{document}

\title{Toward a First-Principles Calculation of Electroweak Box Diagrams}

\author{Chien-Yeah Seng$^{a}$}
\author{Ulf-G. Mei\ss{}ner$^{a,b,c}$}

\affiliation{$^{a}$Helmholtz-Institut f\"ur Strahlen- und Kernphysik and Bethe Center for Theoretical Physics,\\ Universit\"at Bonn,
  53115 Bonn, Germany}
\affiliation{$^{b}$Institute for Advanced Simulation, Institut f\"ur Kernphysik and J\"ulich Center for Hadron Physics,
  Forschungszentrum J\"ulich, 52425 J\"ulich, Germany}
\affiliation{$^{c}$Tbilisi State  University,  0186 Tbilisi, Georgia}

\date{\today}

\begin{abstract}
We derive a Feynman-Hellmann theorem relating the second-order nucleon energy shift resulting from the introduction of periodic source terms
of electromagnetic and isovector axial currents to the parity-odd nucleon structure function $F_3^N$. It is a crucial ingredient in the theoretical
study of the $\gamma W$ and $\gamma Z$ box diagrams that are known to suffer from large hadronic uncertainties. We demonstrate that for a
given $Q^2$, one only needs to compute a small number of energy shifts in order to obtain the required inputs for the box diagrams.
Future lattice calculations based on this approach may shed new light on various topics in precision physics including the
refined determination of the Cabibbo-Kobayashi-Maskawa matrix elements and the weak mixing angle.
\end{abstract}

\maketitle

\begin{figure}
	\begin{centering}
		\includegraphics[width=0.7\linewidth]{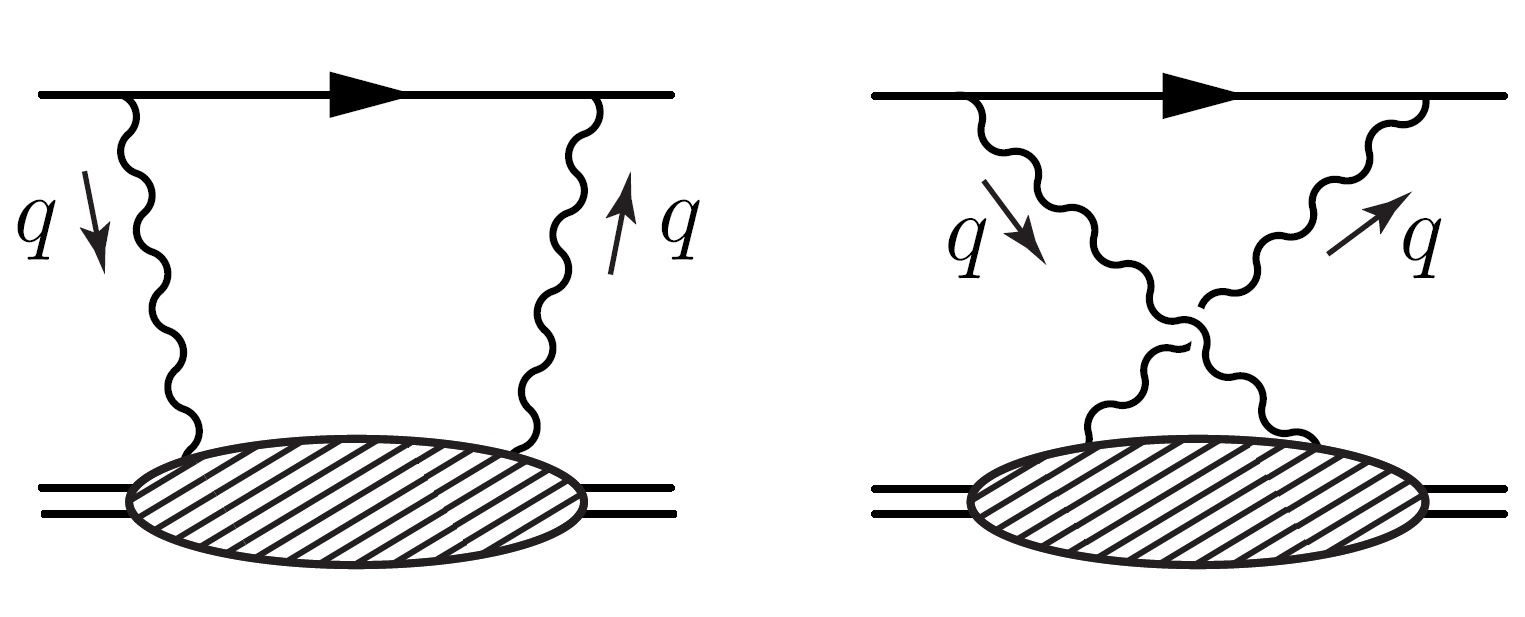}
		\par\end{centering}
	        \caption{Direct and crossed box diagrams. Single and double lines represent a lepton and a hadron, respectively.
                  The blob represents hadronic excitations and the wiggly lines denote the gauge bosons.
                  \label{fig:box}}
\end{figure}

The electroweak box diagrams involving the exchange of a photon and a heavy gauge boson ($W^\pm/Z$) between a lepton and a hadron
(see Fig.~\ref{fig:box}) represent an important component in the standard model (SM) electroweak radiative corrections that enter
various low-energy processes such as semileptonic decays of hadrons and parity-violating lepton-hadron scatterings. These are powerful
tools in extractions of SM weak parameters. The precise calculations of such diagrams are, however, extremely difficult because they
are sensitive to the loop momentum $q$ at all scales and include contributions from all possible virtual hadronic intermediate states
which properties are governed by quantum chromodynamics (QCD) in its nonperturbative regime. Hence, they are one of the main sources
of theoretical uncertainty in the extracted weak parameters such as the Cabibbo-Kobayashi-Maskawa (CKM) matrix
elements~\cite{Hardy:2014qxa,Hardy:2018zsb} and the weak mixing angle~\cite{Kumar:2013yoa} at low scale.

Modern treatments of the box diagrams are based on the pioneering work by Sirlin~\cite{Sirlin:1977sv} in the late 1970s that separates
the diagrams into ``model-independent" and ``model-dependent" terms, of which the former can be reduced to known quantities
by means of current algebra. The model-dependent terms, on the other hand, consist of the interference between the
electromagnetic and the axial weak currents, and are plagued with large hadronic uncertainties at $Q^2\lesssim1$ GeV$^2$. Earlier
attempts to constrain these terms include varying the infrared cutoff~\cite{Marciano:1982mm,Marciano:1983ss,Marciano:1985pd,Bardin:2001ii}
and the use of interpolating functions~\cite{Marciano:2005ec}, but all these methods suffer from nonimprovable theoretical uncertainties. 
The recent introduction of  dispersion relations in treatments of the
$\gamma Z$~\cite{Gorchtein:2008px,Gorchtein:2011mz,Blunden:2011rd,Rislow:2013vta}
and $\gamma W$~\cite{Seng:2018yzq,Seng:2018qru,Gorchtein:2018fxl} boxes provides a better starting point to the problem by
expressing the loop integral in terms of parity-odd structure functions. Since the latter depend on on-shell intermediate hadronic
states, one could in principle relate them to experimental data. Unfortunately, at the hadronic scale such data either do not exist
or belong to a separate isospin channel which can only be related to our desired structure functions within a model. 

First-principles calculations of the parity-odd structure functions from lattice QCD have not yet been thoroughly investigated,
and are expected to be challenging due to the existence of multihadron final states. Moreover, most of the recent developments in the
lattice calculation of parton distribution functions (see, e.g. Ref. \cite{Lin:2017snn}) do not apply here because their applicability
is restricted to large $Q^2$. But at the same time we also observe an encouraging development in the application of the
Feynman-Hellmann theorem (FHT)~\cite{Feynman:1939zza,Hellmann}, where external source terms are added to the Hamiltonian, and the
required hadronic matrix elements of the source operator could be related to the energy shift of the corresponding hadron
which is easier to obtain on lattice as it avoids the calculation of complicated (and potentially noisy) contraction diagrams. A
nonzero momentum transfer $\vec{q}$ can also be introduced by adopting a periodic source term. Such a method shows great potential
in the calculation of hadron electromagnetic form factors~\cite{Chambers:2017tuf}, Compton scattering
amplitude~\cite{Agadjanov:2016cjc,Agadjanov:2018yxh}, parity-even nucleon structure functions~\cite{Chambers:2017dov},
and hadron resonances~\cite{RuizdeElvira:2017aet}. Furthermore, it does not involve any operator product expansion so its applicability
is not restricted to large $Q^2$.

Based on the  developments above, we propose in this Letter a new method to study the $\gamma W$/$\gamma Z$ boxes, namely to
compute a generalized parity-odd forward Compton tensor on the lattice through the second-order nucleon energy shift upon introducing
two periodic source terms, and solve for the moments of $F_3$ through a dispersion relation. We will demonstrate that for a given $Q^2$,
the calculation of a few energy shifts already provides sufficient information about the integrand of the box diagrams, and such a
calculation is completely executable with the computational power in the current lattice community. When data are accumulated for
sufficiently many values of $Q^2$ at the hadronic scale, one will eventually be able to remove the hadronic uncertainties in the
electroweak boxes and provide a satisfactory solution to this long-lasting problem in precision physics.

We start by defining the electromagnetic current and the isovector axial current {\color{black}(here we neglect the strange current just to simplify our discussions of the two examples below, but it is not a necessary approximation) }:
\begin{eqnarray}
J_{em}^\mu&=&(2/3)\bar{u}\gamma^\mu u-(1/3)\bar{d}\gamma^\mu d\nonumber\\
J_A^\mu&=&\bar{u}\gamma^\mu\gamma_5 u-\bar{d}\gamma^\mu\gamma_5 d.
\end{eqnarray}
The spin-independent, parity-odd nucleon structure function $F_3^N$  ($N=p,n$) can be defined through the hadronic tensor:
\begin{eqnarray}
W^{\mu\nu}_N(p,q)&=&\frac{1}{4\pi}\int d^4xe^{iq\cdot x}\left\langle N(\vec{p})\right|[J_{em}^\mu(x),J_A^\nu(0)]\left|N(\vec{p})
\right\rangle\nonumber\\
&=&-\frac{i\varepsilon^{\mu\nu\alpha\beta}q_\alpha p_\beta}{2p\cdot q}F_3^N(x_B,Q^2),
\end{eqnarray}
where $x_B=Q^2/(2p\cdot q)$ is the Bjorken variable which lies between $-1$ and $1$, and $\varepsilon^{0123}=-1$. We stress that it is more natural to include negative values of $x_B$ because in a dispersion relation involving $F_3^N(x_B,Q^2)$, $x_B$ acts as the integration variable in the Cauchy integral that could lie on both the positive and the negative real axes. Notice that the spin
label in the nucleon states are suppressed for simplicity. From that we may define the so-called ``first Nachtmann moment" of
$F_3^N$ as~\cite{Nachtmann:1973mr,Nachtmann:1974aj}
\begin{equation}
M_1[F_3^{N}]=\int_0^1dx\Pi(x,Q^2)F_3^{N}(x,Q^2),
\end{equation}
where
 \begin{equation}
\Pi(x,Q^2)=\frac{4}{3}\frac{1+2\sqrt{1+4m_N^2x^2/Q^2}}{(1+\sqrt{1+4m_N^2x^2/Q^2})^2}
\end{equation}
and $m_N$ is the nucleon mass.

To see the physical relevance of the definitions above, we shall briefly discuss some recent progress in the study of two weak
processes that play central roles in low-energy precision tests of SM:

First, superallowed nuclear $\beta$ decays represent the best avenues for the measurement of the CKM matrix element $V_{ud}$
as the corresponding weak nuclear matrix element is protected at tree level by the conserved vector current. With the inclusion of
higher-order corrections one obtains~\cite{Hardy:2014qxa}:
\begin{equation}
|V_{ud}|^2=\frac{2984.432(3)\:\mathrm{s}}{\mathcal{F}t(1+\Delta_R^V)}
\end{equation}
where $\mathcal{F}t$ is {\color{black}the product between the half-life $t$ and the statistical function $f$, but} modified by nuclear-dependent corrections. $\Delta_R^V$ represents the
nucleus-independent radiative correction. The main theoretical uncertainty of $|V_{ud}|$ comes from $\Delta_R^V$, which in turn
acquires its largest uncertainty from the interference between the isosinglet electromagnetic current and the axial charged
weak current in the $\gamma W$ box diagram. The latter can be expressed as:
\begin{equation}
\left(\Delta_R^V\right)_{\gamma W}^{VA}=\int_0^\infty\frac{dQ^2}{Q^2}\frac{3\alpha}{\pi}\frac{M_W^2}{M_W^2+Q^2}M_1[F_3^{(0)}],
\end{equation}
where $F_3^{(0)}=-(1/4)(F_3^p-F_3^n)$ through isospin symmetry.
A recent determination of $\Delta_R^V$ based on a dispersion relation and neutrino scattering data gives 0.02467(22)~\cite{Seng:2018yzq},
which lies significantly above the previous sate-of-the-art result of 0.02361(38)~\cite{Marciano:2005ec} and leads to an apparent violation
of the first-row CKM unitarity at the level of 4$\sigma$ that calls for an immediate resolution. Besides, scrutinizing the problems
in $V_{ud}$ will also lead to a better determination of $V_{us}$, because one of the main measuring channels of the latter, the
$K\to\mu\nu(\gamma)$ decay, probes the ratio $|V_{us}|/|V_{ud}|$.

Second, we look at parity-odd $ep$ scattering. The measurement of the proton weak charge $Q_W^p$ in the almost-forward elastic $ep$ scattering
is a powerful probe of the physics beyond SM due to the accidental suppression of its tree-level value $1-4\sin^2\theta_W$,
with $\theta_W$  the weak mixing angle. After including one-loop electroweak radiative corrections, the quantity reads~\cite{Erler:2003yk}
\begin{eqnarray}
Q_W^p&=&(1+\Delta \rho+\Delta_e)[1-4\sin^2\theta_W(0)+\Delta_e']\nonumber\\
&&+\Box_{WW}+\Box_{ZZ}+\Box_{\gamma Z},
\end{eqnarray} 
among which $\Box_{\gamma Z}$ represents the contribution from the $\gamma Z$ box that bears the largest hadronic uncertainty. In the limit
of vanishing beam energy, it takes the following form:
\begin{equation}
\Box_{\gamma Z}=\int_0^\infty \frac{dQ^2}{Q^2}\frac{3\alpha}{2\pi}v_e\frac{M_Z^2}{M_Z^2+Q^2}M_1[F_3^{\gamma Z}],
\end{equation}
where $v_e$ is the electron weak charge and $F_3^{\gamma Z}=-F_3^p$. {\color{black}A recent} estimation of $\Box_{\gamma Z}$ reads
$0.0044(4)$~\cite{Blunden:2011rd}. In view of the upcoming P2 experiment at the Mainz Energy-Recovering Superconducting Accelerator (MESA) that aims for the measurement of $\sin^2\theta_W$
to a precision of $0.15\%$~\cite{Becker:2018ggl}, it is necessary for a revisit of the $\gamma Z$ box to proceed coherently with $\gamma W$ in order to ensure there is no unaccounted systematics as recently discovered in the latter \cite{Seng:2018yzq,Seng:2018qru}.

From the two examples above one sees that the object of interest is the first Nachtmann moment of $F_3^N$, which probes different
on-shell intermediate states at different $Q^2$. The analysis of the data accumulated for an analogous parity-odd structure function
$F_3^{WW}$ resulting from the interference between the vector and axial charged weak current in inclusive $\nu p/\bar{\nu}p$
scattering indicates that (1) at $Q^2< 0.1$ GeV$^2$ the first Nachtmann moment is saturated by
the contribution from the elastic intermediate state and the lowest nucleon resonances~\cite{Bolognese:1982zd} of which sufficient
data are available, and (2) at $Q^2>2$ GeV$^2$ it is well described by a parton model with well-known perturbative QCD
corrections~\cite{Kataev:1994ty,Kim:1998kia} {\color{black}(see also, Sec.~IV of Ref.~\cite{Seng:2018qru} for a detailed description of the dominant physics that takes place at different $Q^2$)}. On the other hand, multihadron intermediate states dominate at $Q^2\lesssim 1$ GeV$^2$,
and a first-principles theoretical description at this region is absent so far. Although there are attempts to relate, say, $F_3^{(0)}$
to the measured $F_3^{WW}$ in this region, such a relation is only established within a model because it belongs to different
isospin channels. Therefore, the goal of this Letter is to outline a method that allows for a reliable first-principles
calculation of $M_1[F_3^N]$ at $Q^2\lesssim 1$ GeV$^2$. 

To achieve this goal, we consider the following generalized forward Compton tensor:
\begin{eqnarray}
  T_N^{\mu\nu}(p,q)&=&\int d^4xe^{iq\cdot x}\left\langle N(\vec{p})\right|T\left\{J_{em}^\mu(x)J_A^\nu(0)\right\}
  \left|N(\vec{p})\right\rangle\nonumber\\
  &=&-\frac{i\varepsilon^{\mu\nu\alpha\beta}q_\alpha p_\beta}{2p\cdot q}T_3^N(\omega,Q^2),\label{eq:Compton}
\end{eqnarray}
where $\omega=1/x_B=2p\cdot q/Q^2$, and time-reversal invariance requires $T_3^N(\omega,Q^2)$ to be an odd function of $\omega$.
Unlike the structure function, here we do not require the intermediate states to stay on shell, so one could have $|\omega|<1$.
A dispersion relation exists between $T_3^N$ and $F_3^N$:
\begin{equation}
T_3^N(\omega,Q^2)=-4i\omega\int_0^1dx\frac{F_3^N(x,Q^2)}{1-\omega^2x^2}.	\label{eq:dispersion}
\end{equation}
Therefore, if one is able to compute $T_3^N(\omega,Q^2)$ at several points of $\omega$ below the elastic threshold, then one could
extract useful information about the structure function $F_3^N$ through Eq.~\eqref{eq:dispersion}.

Our approach is to make use of the second-order FHT that relates the second derivative of the nucleon energy upon the introduction
of periodic source terms to $T_3^N$ below threshold. Let us first state our result here. We define the momentum transfer
$q^\mu=(0,q_x,q_y,q_z)$ so that $Q^2=\vec{q}^2$ and $\omega=-2\vec{p}\cdot\vec{q}/\vec{q}^2$, and throughout this work we
impose the off-shell condition, i.e. $|\omega|=2|\vec{p}\cdot\vec{q}|/\vec{q}^2<1$. We consider the addition of two external
source terms to the ordinary QCD Hamiltonian (we choose $\mu=2$ and $\nu=3$ to be definite):
{\color{black}
\begin{eqnarray}
H_\lambda(t)&=&H_0(t)+2\lambda_1\int d^3x\cos(\vec{q}\cdot\vec{x})J_{em}^2(\vec{x},t)\nonumber\\
&&-2\lambda_2\int d^3x\sin(\vec{q}\cdot\vec{x})J_A^3(\vec{x},t).\label{eq:Hlambda}
\end{eqnarray}
}
The unperturbed nucleon energy with momentum $\vec{p}$ is simply $E_N(\vec{p})=\sqrt{m_N^2+\vec{p}^2}$. After the introduction of the
external source terms, this energy becomes $E_{N,\lambda}(\vec{p})$. We remind the readers that, since the source terms break translational
symmetry, the nucleon eigenstate with energy $E_{N,\lambda}(\vec{p})$ is no longer a momentum eigenstate. The second-order FHT states that:
\begin{equation}
\left(\frac{\partial^2E_{N,\lambda}(\vec{p})}{\partial\lambda_1\partial\lambda_2}\right)_{\lambda=0}=\frac{iq_x}{Q^2\omega}
T_3^N(\omega,Q^2).\label{eq:FHtheorem}
\end{equation}
One could then express the amplitude $T_3^N$ in terms of the dispersion integral~\eqref{eq:dispersion} to obtain
\begin{equation}
  \left(\frac{\partial^2E_{N,\lambda}(\vec{p})}{\partial\lambda_1\partial\lambda_2}\right)_{\lambda=0}=\frac{4q_x}{Q^2}
  \int_0^1dx\frac{F_3^N(x,Q^2)}{1-\omega^2x^2}~,\label{eq:central}
\end{equation}
which is the central result of this Letter. For later convenience, we define the function $\Lambda(x,\omega)=1/(1-\omega^2x^2)$.

Below we shall outline a proof of Eq.~\eqref{eq:FHtheorem} based on the Euclidean path integral, which is closely connected to
standard treatments in lattice QCD~\cite{Bouchard:2016heu} {\color{black}(we also refer interested readers to Ref. \cite{Agadjanov:2018yxh} that contains all details of an almost identical derivation for the case of the parity-even Compton amplitude)}. Throughout, Euclidean quantities will be labeled by a subscript
$\mathbb{E}$. Also, if a quantity is supposed to be affected by the source terms but appears without a subscript $\lambda$, that implies
its limit at $\lambda_1,\lambda_2\rightarrow 0$. First, the existence of extra source terms in Eq.~\eqref{eq:Hlambda} implies a
shift of the Euclidean action:
\begin{eqnarray}
S_{\mathbb{E},\lambda}&=&S^0_\mathbb{E}+2\lambda_1\int d^4x_\mathbb{E}\cos(\vec{q}\cdot\vec{x})J_{em}^2(x_\mathbb{E})\nonumber\\
&&-2\lambda_2\int d^4x_\mathbb{E}\sin(\vec{q}\cdot\vec{x})J_A^3(x_\mathbb{E}).
\end{eqnarray}
Next, we define a two-point correlation function:
\begin{equation}
C_\lambda^N(\vec{p},t_\mathbb{E})=\int d^3xe^{-i\vec{p}\cdot\vec{x}}\left\langle\Omega_\lambda\right|T\{\chi_N(\vec{x},t_\mathbb{E})\chi_N^\dagger(0)\}\left|\Omega_\lambda\right\rangle,
\end{equation}
with $t_\mathbb{E}>0$. Here, $\chi_N$ is an interpolating operator that possesses the same quantum numbers as the nucleon $N$. We remind
the readers that a time-ordered correlation function {\color{black} of arbitrary operators $O_i$} with respect to the {\color{black}vacuum state $\left|\Omega_\lambda\right\rangle$} can be expressed in terms of a Euclidean path integral:
\begin{eqnarray}
&&\left\langle\Omega_\lambda\right|T\{O_1(t_{1\mathbb{E}})...O_n(t_{n\mathbb{E}})\}\left|\Omega_\lambda\right\rangle\nonumber\\
&=&\frac{1}{Z_{\mathbb{E},\lambda}}\int D\phi\:O_1(t_{1\mathbb{E}})...O_n(t_{n\mathbb{E}})e^{-S_{\mathbb{E},\lambda}},
\end{eqnarray}
{\color{black}with $Z_{\mathbb{E},\lambda}$ the Euclidean partition function.} Based on the asymptotic behavior of $C_\lambda^N$, we  define an ``effective energy":
\begin{equation}
  E^{\rm eff}_{N,\lambda}(\vec{p};t_\mathbb{E},\tau_\mathbb{E})=\frac{1}{\tau_\mathbb{E}}\ln\left(
  \frac{C_\lambda^N(\vec{p},t_\mathbb{E})}{C_\lambda^N(\vec{p},t_\mathbb{E}+\tau_\mathbb{E})}\right),
\end{equation}
that reduces to the nucleon energy in the large-time limit (which is only true when $|\omega|<1$):
\begin{equation}
\lim_{t_\mathbb{E}\rightarrow \infty} E^{\rm eff}_{N,\lambda}(\vec{p};t_\mathbb{E},\tau_\mathbb{E})=E_{N,\lambda}(\vec{p}).
\end{equation}
Therefore, one may obtain the partial derivatives of $E_{N,\lambda}(\vec{p})$ with respect to $\lambda_i$ through the partial
derivatives of $E_{N,\lambda}^{\rm eff}$. An advantage in doing so is that one could see explicitly that the
``vacuum matrix elements", i.e. terms with $\partial Z_{\mathbb{E},\lambda}/\partial\lambda_i$, do not contribute. We find
that the first derivative vanishes:
\begin{equation}
\left(\frac{\partial E_{N,\lambda}(\vec{p})}{\partial\lambda_i}\right)_{\lambda=0}=\lim_{t_\mathbb{E}\rightarrow 0}\left(\frac{\partial E_{N,\lambda}^{\rm eff}(\vec{p};t_\mathbb{E},\tau_\mathbb{E})}{\partial\lambda_i}\right)_{\lambda=0}=0.
\end{equation}
The underlying reason is simple: the external source terms induce a momentum shift of $\pm \vec{q}$ upon each insertion; therefore,
according to usual perturbation theory, the linear energy shift is proportional to $
\left\langle\vec{p}\right|\left.\vec{p}\pm\vec{q}\right\rangle =0$ for $\vec{q}\neq 0$. 

We are interested in the second derivative of $E_{N,\lambda}(\vec{p})$ which reads
\begin{equation}
  \left(\frac{\partial^2E_{N,\lambda}(\vec{p})}{\partial\lambda_1\partial\lambda_2}\right)_{\lambda=0}\!\!\!\!\!\!=\lim_{t_\mathbb{E}\rightarrow \infty}
  \frac{1}{\tau_\mathbb{E}}\left[\frac{R(\vec{p},\vec{q},t_\mathbb{E})}{C^N(\vec{p},t_\mathbb{E})}-(t_\mathbb{E}\rightarrow t_\mathbb{E}
    +\tau_\mathbb{E})\right]
\end{equation}
where
\begin{eqnarray}
&&R(\vec{p},\vec{q},t_\mathbb{E})=\int d^3xe^{-i\vec{p}\cdot\vec{x}}\times\nonumber\\
&&\!\!\!\!\!\!\!\!\left\langle\Omega\right|T\left\{\chi_N(\vec{x},t_\mathbb{E})\chi_N^\dagger(0)\left(\frac{\partial
S_{\mathbb{E},\lambda}}{\partial\lambda_1}\right)\left(\frac{\partial S_{\mathbb{E},\lambda}}{\partial\lambda_2}\right)\right\}
\left|\Omega\right\rangle~.
\end{eqnarray}
One then splits the time-ordered product in $R(\vec{p},\vec{q},t_\mathbb{E})$ into different time regions, and finds that at large
$t_\mathbb{E}$ the dominant piece is the one with the two currents sandwiched between $\chi_N$ and $\chi_N^\dagger$. We may then
insert two complete sets of states between the interpolating operators and the current product, and since the off-shell condition
ensures $E_N(\vec{p}\pm 2\vec{q})>E_N(\vec{p})$, we find that the dominant piece consists of a time-ordered nucleon matrix element
with the same momentum $\vec{p}$ in the initial and final states. We therefore isolate this piece and make use of the following identity:
\begin{multline}
\int_0^{t_\mathbb{E}}dy^4_\mathbb{E}\int_0^{t_\mathbb{E}}dz^4_\mathbb{E}\left\langle N(\vec{p})\right|T\{J_{em}^2(\vec{y},y_\mathbb{E}^4-z_\mathbb{E}^4)J_A^3(0)\}
\left|N(\vec{p})\right\rangle\\
\rightarrow t_\mathbb{E}\int_{-\infty}^{\infty}dy^4_\mathbb{E}\left\langle N(\vec{p})\right|T\{J_{em}^2(\vec{y},y_\mathbb{E}^4)J_A^3(0)\}
\left|N(\vec{p})\right\rangle
\end{multline}
to obtain:
\begin{eqnarray}
&&E_N(\vec{p})\left(\frac{\partial^2E_{N,\lambda}(\vec{p})}{\partial\lambda_1\partial\lambda_2}\right)_{\lambda=0}=-\int d^4y_\mathbb{E}
\sin(\vec{q}\cdot\vec{y})\times\nonumber\\
&&\left\langle N(\vec{p})\right|T\{J_{em}^2(y_\mathbb{E})J_A^3(0)\}\left|N(\vec{p})\right\rangle.
\end{eqnarray}
We can now switch back to the Minkowskian space time through a Wick rotation: $\int d^4y_\mathbb{E}\to i\int d^4y $. Finally, we substitute
the result into Eq.~\eqref{eq:Compton} and make use of crossing symmetry $T_3(-\omega,Q^2)=-T_3(\omega,Q^2)$ to arrive at Eq.~\eqref{eq:FHtheorem}. This completes the proof.

Now let us discuss the practical use of Eq.~\eqref{eq:central}. Ideally, it allows for a reconstruction of the full structure
function $F_3^N(x,Q^2)$ by calculating the second-order energy shift at $n\gg 1$ discrete points of $\omega$: we simply discretize
the dispersion integral to obtain a matrix equation:
\begin{equation}
\left(\frac{\partial^2E_{N,\lambda}(\vec{p})}{\partial\lambda_1\partial\lambda_2}\right)_{\lambda=0}(\omega_i,Q^2)\approx \sum_{j=1}^n A_{\omega_i,x_j}F_3^N(x_j,Q^2),\label{eq:matrixeq}
\end{equation}
and notice that the matrix $A$ does not possess any singularity with $\omega$ below the elastic threshold. We may then invert $A$ to
obtain $F_3^N(x,Q^2)$ at the discrete points $\{x_j\}$. However, such an approach is accurate only when $n$ is large, which is difficult
to achieve with the current lattice computational power when $Q^2\lesssim 1$ GeV$^2$. To see this, one first recalls that any momentum in
a finite lattice can only take discrete values:
$\vec{k}=(2\pi/L)(n_{kx},n_{ky},n_{kz})$, with $L$ is the spatial lattice size and $\{n_{kx},n_{ky},n_{kz}\}$ are integers. The requirements
that $Q^2=\vec{q}^2\lesssim 1$ GeV$^2$ and $|\omega|=2|\vec{p}\cdot\vec{q}|/\vec{q}^2<1$ imply two conditions:
\begin{eqnarray}
\frac{4\pi^2}{L^2}(n_{qx}^2+n_{qy}^2+n_{qz}^2)&\lesssim &1\:\mathrm{GeV}^2\\
\frac{2|n_{px}n_{qx}+n_{py}n_{qy}+n_{pz}n_{qz}|}{n_{qx}^2+n_{qy}^2+n_{qz}^2}&<&1.\label{eq:condition2}
\end{eqnarray}
In particular, with a fixed choice of $\vec{q}$, the second condition determines the allowed discrete values of $\omega$ at which
the nucleon energy can be extracted on lattice. To understand how low in $Q^2$ one can probe, we consider a typical lattice setup: the configuration cA2.09.48 from the
ETM Collaboration that features a spatial lattice size of $48\times 0.0931~\mathrm{fm}\approx 4.47~\mathrm{fm}$~\cite{Liu:2016cba}.
For such a configuration, we get $Q^2\approx 0.38$~GeV$^2$ with the choice $\vec{q}=(2\pi/L)(2,1,0)$, but Eq.~\eqref{eq:condition2} restricts
the number of allowed $|\omega|$ to three: 0, 2/5, and 4/5. Such a small amount is obviously insufficient to perform the matrix
inversion of Eq.~\eqref{eq:matrixeq} to any satisfactory level of accuracy.

\begin{figure}
	\begin{centering}
		\includegraphics[width=0.8\linewidth]{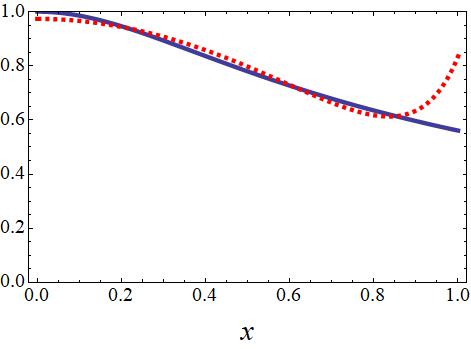}
		\par\end{centering}
	\caption{Comparison between the function $\Pi(x,Q^2)$ at $Q^2=0.38$ GeV$^2$ (blue solid line) and $\Lambda_\mathrm{tot}(x)$ (red dashed line) defined in Eq. \eqref{eq:Lambdatot}.
		\label{fig:fit}}
\end{figure}

Fortunately, in studies of the electroweak boxes we do not need the full $F_3^N(x,Q^2)$ as a function of $x$, but rather its first
Nachtmann moment. Therefore, the real question is whether one could form a linear combination of the functions $\{\Lambda(x,\omega_i)\}$ that appear in the dispersion integral~\eqref{eq:central} with all allowed values of $\omega_i$ to approximate the function $\Pi(x,Q^2)$ to a satisfactory level, especially at small $x$ {\color{black}(because apart from the known, isolated elastic contribution at $x=1$, $F_3^N(x,Q^2)$ is non-zero only at $x<x_\pi=Q^2/(2m_NM_\pi+M_\pi^2+Q^2)$, with $M_\pi$ the pion mass)}. As a proof of principle, let us still consider the example above. We define the following linear combination:
\begin{equation}
\Lambda_\mathrm{tot}(x)=a\Lambda(x,0)+b\Lambda(x,2/5)+c\Lambda(x,4/5),\label{eq:Lambdatot}
\end{equation}	
and fit the parameters $\{a,b,c\}$ to match $\Pi(x,Q^2)$ at $Q^2\approx 0.38$ GeV$^2$. We find that they come to a good agreement at $x<0.9$ with the choice $a=7.82446$, $b=-7.58605$ and $c=0.734787$, as shown in Fig. \ref{fig:fit}. That means we could obtain a very good approximation to $M_1[F_3^N]$ by adding the values of $(Q^2/4q_x)\times(\partial_{\lambda_1}\partial_{\lambda_2}E_{N,\lambda})_{\lambda=0}$ calculated at $\omega=0,2/5,4/5$ with the weighting coefficients $\{a,b,c\}$ respectively. We shall also discuss the efficiency of this procedure for different values of $Q^2$: with the same $L$, at larger $Q^2$ one has more available values of $\omega$ and the global fitting to $\Pi(x,Q^2)$ will be even better; this is encouraging because $Q^2>0.38$ GeV$^2$ already fully covers the so-called ``intermediate distances" in Ref.~\cite{Marciano:2005ec} that contain most of the hadronic uncertainties. On the other hand, at smaller $Q^2$ (such as $Q^2$=0.1~GeV$^2$) the allowed values of $\omega$ are less so one is not able to reproduce $M_1[F_3^N]$ with the same accuracy. The readers, however, should not be discouraged because (1) $\omega=0$ is always an accessible point, which gives the first {\it Mellin} moment of $F_3^N$ according to Eq.~\eqref{eq:central}. This will provide important constraints for model parameterizations of the residual multihadron contributions to $F_3^N(x,Q^2)$ at small $Q^2$, and (2) future efforts in the increase of the lattice size (see, e.g. Ref.~\cite{Luscher:2017cjh}) will then allow for a precise calculation of $M_1[F_3^N]$ at smaller $Q^2$ with our proposed method.

We shall end by commenting on the required level of precision for lattice calculations. We take the $\gamma W$ box as an example:
in Ref.~\cite{Seng:2018qru}, the contribution from multiparticle intermediate states at {\color{black}$Q^2\sim 1$ GeV$^2$} to $\Delta_R^V$ is estimated
to be $(\alpha/\pi)\times(0.48\pm0.07)$ through a simple Regge-exchange model, with a $\sim 15\%$ error coming from the
$\nu p/\bar{\nu}p$ scattering data. Possible systematic errors due to the simplicity of the model itself are not accounted for.
In this sense, a successful lattice calculation of the second-order nucleon energy shift at a few points of $\omega$ with a precision
level of $15\%$ will already be able to match the {\it{precision}} of the model and start to challenge its {\it{accuracy}}. {\color{black}This is completely executable with current lattice techniques as a similar calculation for parity-even structure functions has already been performed in Ref.~\cite{Chambers:2017dov} with a 10\% overall projected error}. Also, by employing a larger $L$ one is able to probe
smaller $Q^2$, and when sufficiently many points of $M_1[F_3^N]$ between $0.1$ GeV$^2<Q^2<1$ GeV$^2$ are determined, one will
be able to reduce the hadronic uncertainties in the $\gamma W$ and $\gamma Z$ boxes to a level compatible with current and future
precision experiments.

\vspace{0.5cm}

\noindent {\bf Acknowledgements} --
The authors thank Akaki Rusetsky, Gerrit Schierholz and Mikhail Gorchtein for inspiring discussions.
This work is supported in part by  the DFG (Grant No. TRR110)
and the NSFC (Grant No. 11621131001) through the funds provided
to the Sino-German CRC 110 ``Symmetries and the Emergence of
Structure in QCD", by the Alexander von Humboldt Foundation through a  Humboldt Research Fellowship (CYS),
by the Chinese  Academy of Sciences (CAS) President's International Fellowship Initiative (PIFI)
(grant no. 2018DM0034) (UGM) and by VolkswagenStiftung (grant no. 93562) (UGM).


\begin{thebibliography}{31}
	\expandafter\ifx\csname natexlab\endcsname\relax\def\natexlab#1{#1}\fi
	\expandafter\ifx\csname bibnamefont\endcsname\relax
	\def\bibnamefont#1{#1}\fi
	\expandafter\ifx\csname bibfnamefont\endcsname\relax
	\def\bibfnamefont#1{#1}\fi
	\expandafter\ifx\csname citenamefont\endcsname\relax
	\def\citenamefont#1{#1}\fi
	\expandafter\ifx\csname url\endcsname\relax
	\def\url#1{\texttt{#1}}\fi
	\expandafter\ifx\csname urlprefix\endcsname\relax\def\urlprefix{URL }\fi
	\providecommand{\bibinfo}[2]{#2}
	\providecommand{\eprint}[2][]{\url{#2}}
	
	\bibitem[{\citenamefont{Hardy and Towner}(2015)}]{Hardy:2014qxa}
	\bibinfo{author}{\bibfnamefont{J.~C.} \bibnamefont{Hardy}} \bibnamefont{and}
	\bibinfo{author}{\bibfnamefont{I.~S.} \bibnamefont{Towner}},
	\bibinfo{journal}{Phys. Rev.} \textbf{\bibinfo{volume}{C91}},
	\bibinfo{pages}{025501} (\bibinfo{year}{2015}), \eprint{1411.5987}.
	
	\bibitem[{\citenamefont{Hardy and Towner}(2018)}]{Hardy:2018zsb}
	\bibinfo{author}{\bibfnamefont{J.~C.} \bibnamefont{Hardy}} \bibnamefont{and}
	\bibinfo{author}{\bibfnamefont{I.~S.} \bibnamefont{Towner}}, in
	\emph{\bibinfo{booktitle}{{13th Conference on the Intersections of Particle
				and Nuclear Physics (CIPANP 2018) Palm Springs, California, USA, May 29-June
				3, 2018}}} (\bibinfo{year}{2018}), \eprint{1807.01146}.
	
	\bibitem[{\citenamefont{Kumar et~al.}(2013)\citenamefont{Kumar, Mantry,
			Marciano, and Souder}}]{Kumar:2013yoa}
	\bibinfo{author}{\bibfnamefont{K.~S.} \bibnamefont{Kumar}},
	\bibinfo{author}{\bibfnamefont{S.}~\bibnamefont{Mantry}},
	\bibinfo{author}{\bibfnamefont{W.~J.} \bibnamefont{Marciano}},
	\bibnamefont{and} \bibinfo{author}{\bibfnamefont{P.~A.}
		\bibnamefont{Souder}}, \bibinfo{journal}{Ann. Rev. Nucl. Part. Sci.}
	\textbf{\bibinfo{volume}{63}}, \bibinfo{pages}{237} (\bibinfo{year}{2013}),
	\eprint{1302.6263}.
	
	\bibitem[{\citenamefont{Sirlin}(1978)}]{Sirlin:1977sv}
	\bibinfo{author}{\bibfnamefont{A.}~\bibnamefont{Sirlin}},
	\bibinfo{journal}{Rev. Mod. Phys.} \textbf{\bibinfo{volume}{50}},
	\bibinfo{pages}{573} (\bibinfo{year}{1978}), \bibinfo{note}{[Erratum: Rev.
		Mod. Phys. {\bf 50}, 905 (1978)]}.
	
	\bibitem[{\citenamefont{Marciano and Sirlin}(1983)}]{Marciano:1982mm}
	\bibinfo{author}{\bibfnamefont{W.~J.} \bibnamefont{Marciano}} \bibnamefont{and}
	\bibinfo{author}{\bibfnamefont{A.}~\bibnamefont{Sirlin}},
	\bibinfo{journal}{Phys. Rev.} \textbf{\bibinfo{volume}{D27}},
	\bibinfo{pages}{552} (\bibinfo{year}{1983}).
	
	\bibitem[{\citenamefont{Marciano and Sirlin}(1984)}]{Marciano:1983ss}
	\bibinfo{author}{\bibfnamefont{W.~J.} \bibnamefont{Marciano}} \bibnamefont{and}
	\bibinfo{author}{\bibfnamefont{A.}~\bibnamefont{Sirlin}},
	\bibinfo{journal}{Phys. Rev.} \textbf{\bibinfo{volume}{D29}},
	\bibinfo{pages}{75} (\bibinfo{year}{1984}), \bibinfo{note}{[Erratum: Phys.
		Rev. {\bf D31}, 213 (1985)]}.
	
	\bibitem[{\citenamefont{Marciano and Sirlin}(1986)}]{Marciano:1985pd}
	\bibinfo{author}{\bibfnamefont{W.~J.} \bibnamefont{Marciano}} \bibnamefont{and}
	\bibinfo{author}{\bibfnamefont{A.}~\bibnamefont{Sirlin}},
	\bibinfo{journal}{Phys. Rev. Lett.} \textbf{\bibinfo{volume}{56}},
	\bibinfo{pages}{22} (\bibinfo{year}{1986}).
	
	\bibitem[{\citenamefont{Bardin et~al.}(2001)\citenamefont{Bardin, Christova,
			Kalinovskaya, and Passarino}}]{Bardin:2001ii}
	\bibinfo{author}{\bibfnamefont{D.~{\relax Yu}.} \bibnamefont{Bardin}},
	\bibinfo{author}{\bibfnamefont{P.}~\bibnamefont{Christova}},
	\bibinfo{author}{\bibfnamefont{L.}~\bibnamefont{Kalinovskaya}},
	\bibnamefont{and}
	\bibinfo{author}{\bibfnamefont{G.}~\bibnamefont{Passarino}},
	\bibinfo{journal}{Eur. Phys. J.} \textbf{\bibinfo{volume}{C22}},
	\bibinfo{pages}{99} (\bibinfo{year}{2001}), \eprint{hep-ph/0102233}.
	
	\bibitem[{\citenamefont{Marciano and Sirlin}(2006)}]{Marciano:2005ec}
	\bibinfo{author}{\bibfnamefont{W.~J.} \bibnamefont{Marciano}} \bibnamefont{and}
	\bibinfo{author}{\bibfnamefont{A.}~\bibnamefont{Sirlin}},
	\bibinfo{journal}{Phys. Rev. Lett.} \textbf{\bibinfo{volume}{96}},
	\bibinfo{pages}{032002} (\bibinfo{year}{2006}), \eprint{hep-ph/0510099}.
	
	\bibitem[{\citenamefont{Gorchtein and Horowitz}(2009)}]{Gorchtein:2008px}
	\bibinfo{author}{\bibfnamefont{M.}~\bibnamefont{Gorchtein}} \bibnamefont{and}
	\bibinfo{author}{\bibfnamefont{C.~J.} \bibnamefont{Horowitz}},
	\bibinfo{journal}{Phys. Rev. Lett.} \textbf{\bibinfo{volume}{102}},
	\bibinfo{pages}{091806} (\bibinfo{year}{2009}), \eprint{0811.0614}.
	
	\bibitem[{\citenamefont{Gorchtein et~al.}(2011)\citenamefont{Gorchtein,
			Horowitz, and Ramsey-Musolf}}]{Gorchtein:2011mz}
	\bibinfo{author}{\bibfnamefont{M.}~\bibnamefont{Gorchtein}},
	\bibinfo{author}{\bibfnamefont{C.~J.} \bibnamefont{Horowitz}},
	\bibnamefont{and} \bibinfo{author}{\bibfnamefont{M.~J.}
		\bibnamefont{Ramsey-Musolf}}, \bibinfo{journal}{Phys. Rev.}
	\textbf{\bibinfo{volume}{C84}}, \bibinfo{pages}{015502}
	(\bibinfo{year}{2011}), \eprint{1102.3910}.
	
	\bibitem[{\citenamefont{Blunden et~al.}(2011)\citenamefont{Blunden,
			Melnitchouk, and Thomas}}]{Blunden:2011rd}
	\bibinfo{author}{\bibfnamefont{P.~G.} \bibnamefont{Blunden}},
	\bibinfo{author}{\bibfnamefont{W.}~\bibnamefont{Melnitchouk}},
	\bibnamefont{and} \bibinfo{author}{\bibfnamefont{A.~W.}
		\bibnamefont{Thomas}}, \bibinfo{journal}{Phys. Rev. Lett.}
	\textbf{\bibinfo{volume}{107}}, \bibinfo{pages}{081801}
	(\bibinfo{year}{2011}), \eprint{1102.5334}.
	
	\bibitem[{\citenamefont{Rislow and Carlson}(2013)}]{Rislow:2013vta}
	\bibinfo{author}{\bibfnamefont{B.~C.} \bibnamefont{Rislow}} \bibnamefont{and}
	\bibinfo{author}{\bibfnamefont{C.~E.} \bibnamefont{Carlson}},
	\bibinfo{journal}{Phys. Rev.} \textbf{\bibinfo{volume}{D88}},
	\bibinfo{pages}{013018} (\bibinfo{year}{2013}), \eprint{1304.8113}.
	
	\bibitem[{\citenamefont{Seng et~al.}(2018{\natexlab{a}})\citenamefont{Seng,
			Gorchtein, Patel, and Ramsey-Musolf}}]{Seng:2018yzq}
	\bibinfo{author}{\bibfnamefont{C.-Y.} \bibnamefont{Seng}},
	\bibinfo{author}{\bibfnamefont{M.}~\bibnamefont{Gorchtein}},
	\bibinfo{author}{\bibfnamefont{H.~H.} \bibnamefont{Patel}}, \bibnamefont{and}
	\bibinfo{author}{\bibfnamefont{M.~J.} \bibnamefont{Ramsey-Musolf}},
	\bibinfo{journal}{Phys. Rev. Lett.} \textbf{\bibinfo{volume}{121}},
	\bibinfo{pages}{241804} (\bibinfo{year}{2018}{\natexlab{a}}),
	\eprint{1807.10197}.
	
	\bibitem[{\citenamefont{Seng et~al.}(2018{\natexlab{b}})\citenamefont{Seng,
			Gorchtein, and Ramsey-Musolf}}]{Seng:2018qru}
	\bibinfo{author}{\bibfnamefont{C.~Y.} \bibnamefont{Seng}},
	\bibinfo{author}{\bibfnamefont{M.}~\bibnamefont{Gorchtein}},
	\bibnamefont{and} \bibinfo{author}{\bibfnamefont{M.~J.}
		\bibnamefont{Ramsey-Musolf}} (\bibinfo{year}{2018}{\natexlab{b}}),
	\eprint{1812.03352}.
	
	\bibitem[{\citenamefont{Gorchtein}(2018)}]{Gorchtein:2018fxl}
	\bibinfo{author}{\bibfnamefont{M.}~\bibnamefont{Gorchtein}}
	(\bibinfo{year}{2018}), \eprint{1812.04229}.
	
	\bibitem[{\citenamefont{Lin et~al.}(2018)}]{Lin:2017snn}
	\bibinfo{author}{\bibfnamefont{H.-W.} \bibnamefont{Lin}} \bibnamefont{et~al.},
	\bibinfo{journal}{Prog. Part. Nucl. Phys.} \textbf{\bibinfo{volume}{100}},
	\bibinfo{pages}{107} (\bibinfo{year}{2018}), \eprint{1711.07916}.

        \bibitem{Feynman:1939zza}
          R.~P.~Feynman,
          Phys.\ Rev.\  {\bf 56}, 340 (1939).

         \bibitem{Hellmann}
          H. Hellmann,  {\em Einf\"uhrung in die Quantenchemie,} Deuticke, Leipzig und Wien (1937).
           
        \bibitem[{\citenamefont{Chambers
			et~al.}(2017{\natexlab{a}})}]{Chambers:2017tuf}
	\bibinfo{author}{\bibfnamefont{A.~J.} \bibnamefont{Chambers}}
	\bibnamefont{et~al.} (\bibinfo{collaboration}{QCDSF, UKQCD, CSSM}),
	\bibinfo{journal}{Phys. Rev.} \textbf{\bibinfo{volume}{D96}},
	\bibinfo{pages}{114509} (\bibinfo{year}{2017}{\natexlab{a}}),
	\eprint{1702.01513}.
	
	\bibitem[{\citenamefont{Agadjanov et~al.}(2017)\citenamefont{Agadjanov,
			Mei\ss{}ner, and Rusetsky}}]{Agadjanov:2016cjc}
	\bibinfo{author}{\bibfnamefont{A.}~\bibnamefont{Agadjanov}},
	\bibinfo{author}{\bibfnamefont{U.-G.} \bibnamefont{Mei\ss{}ner}},
	\bibnamefont{and} \bibinfo{author}{\bibfnamefont{A.}~\bibnamefont{Rusetsky}},
	\bibinfo{journal}{Phys. Rev.} \textbf{\bibinfo{volume}{D95}},
	\bibinfo{pages}{031502} (\bibinfo{year}{2017}), \eprint{1610.05545}.
	
	\bibitem[{\citenamefont{Agadjanov et~al.}(2019)\citenamefont{Agadjanov,
			Mei\ss{}ner, and Rusetsky}}]{Agadjanov:2018yxh}
	\bibinfo{author}{\bibfnamefont{A.}~\bibnamefont{Agadjanov}},
	\bibinfo{author}{\bibfnamefont{U.-G.} \bibnamefont{Mei\ss{}ner}},
	\bibnamefont{and} \bibinfo{author}{\bibfnamefont{A.}~\bibnamefont{Rusetsky}},
	\bibinfo{journal}{Phys. Rev.} \textbf{\bibinfo{volume}{D99}},
	\bibinfo{pages}{054501} (\bibinfo{year}{2019}), \eprint{1812.06013}.
	
	\bibitem[{\citenamefont{Chambers
			et~al.}(2017{\natexlab{b}})\citenamefont{Chambers, Horsley, Nakamura, Perlt,
			Rakow, Schierholz, Schiller, Somfleth, Young, and
			Zanotti}}]{Chambers:2017dov}
	\bibinfo{author}{\bibfnamefont{A.~J.} \bibnamefont{Chambers}},
	\bibinfo{author}{\bibfnamefont{R.}~\bibnamefont{Horsley}},
	\bibinfo{author}{\bibfnamefont{Y.}~\bibnamefont{Nakamura}},
	\bibinfo{author}{\bibfnamefont{H.}~\bibnamefont{Perlt}},
	\bibinfo{author}{\bibfnamefont{P.~E.~L.} \bibnamefont{Rakow}},
	\bibinfo{author}{\bibfnamefont{G.}~\bibnamefont{Schierholz}},
	\bibinfo{author}{\bibfnamefont{A.}~\bibnamefont{Schiller}},
	\bibinfo{author}{\bibfnamefont{K.}~\bibnamefont{Somfleth}},
	\bibinfo{author}{\bibfnamefont{R.~D.} \bibnamefont{Young}}, \bibnamefont{and}
	\bibinfo{author}{\bibfnamefont{J.~M.} \bibnamefont{Zanotti}},
	\bibinfo{journal}{Phys. Rev. Lett.} \textbf{\bibinfo{volume}{118}},
	\bibinfo{pages}{242001} (\bibinfo{year}{2017}{\natexlab{b}}),
	\eprint{1703.01153}.

        \bibitem{RuizdeElvira:2017aet} 
        J.~Ruiz de Elvira, U.-G.~Mei{\ss}ner, A.~Rusetsky and G.~Schierholz,
        Eur.\ Phys.\ J.\ C {\bf 77}, no. 10, 659 (2017).
        
	\bibitem[{\citenamefont{Nachtmann}(1973)}]{Nachtmann:1973mr}
	\bibinfo{author}{\bibfnamefont{O.}~\bibnamefont{Nachtmann}},
	\bibinfo{journal}{Nucl. Phys.} \textbf{\bibinfo{volume}{B63}},
	\bibinfo{pages}{237} (\bibinfo{year}{1973}).
	
	\bibitem[{\citenamefont{Nachtmann}(1974)}]{Nachtmann:1974aj}
	\bibinfo{author}{\bibfnamefont{O.}~\bibnamefont{Nachtmann}},
	\bibinfo{journal}{Nucl. Phys.} \textbf{\bibinfo{volume}{B78}},
	\bibinfo{pages}{455} (\bibinfo{year}{1974}).
	
	\bibitem[{\citenamefont{Erler et~al.}(2003)\citenamefont{Erler, Kurylov, and
			Ramsey-Musolf}}]{Erler:2003yk}
	\bibinfo{author}{\bibfnamefont{J.}~\bibnamefont{Erler}},
	\bibinfo{author}{\bibfnamefont{A.}~\bibnamefont{Kurylov}}, \bibnamefont{and}
	\bibinfo{author}{\bibfnamefont{M.~J.} \bibnamefont{Ramsey-Musolf}},
	\bibinfo{journal}{Phys. Rev.} \textbf{\bibinfo{volume}{D68}},
	\bibinfo{pages}{016006} (\bibinfo{year}{2003}), \eprint{hep-ph/0302149}.
	
	\bibitem[{\citenamefont{Becker et~al.}(2018)}]{Becker:2018ggl}
	\bibinfo{author}{\bibfnamefont{D.}~\bibnamefont{Becker}} \bibnamefont{et~al.},
	\bibinfo{journal}{Eur. Phys. J.} \textbf{\bibinfo{volume}{A54}},
	\bibinfo{pages}{208} (\bibinfo{year}{2018}), \eprint{1802.04759}.
	
	\bibitem[{\citenamefont{Bolognese et~al.}(1983)\citenamefont{Bolognese, Fritze,
			Morfin, Perkins, Powell, and Scott}}]{Bolognese:1982zd}
	\bibinfo{author}{\bibfnamefont{T.}~\bibnamefont{Bolognese}},
	\bibinfo{author}{\bibfnamefont{P.}~\bibnamefont{Fritze}},
	\bibinfo{author}{\bibfnamefont{J.}~\bibnamefont{Morfin}},
	\bibinfo{author}{\bibfnamefont{D.~H.} \bibnamefont{Perkins}},
	\bibinfo{author}{\bibfnamefont{K.}~\bibnamefont{Powell}}, \bibnamefont{and}
	\bibinfo{author}{\bibfnamefont{W.~G.} \bibnamefont{Scott}}
	(\bibinfo{collaboration}{Aachen-Bonn-CERN-Democritos-London-Oxford-Saclay}),
	\bibinfo{journal}{Phys. Rev. Lett.} \textbf{\bibinfo{volume}{50}},
	\bibinfo{pages}{224} (\bibinfo{year}{1983}).
	
	\bibitem[{\citenamefont{Kataev and Sidorov}(1994)}]{Kataev:1994ty}
	\bibinfo{author}{\bibfnamefont{A.~L.} \bibnamefont{Kataev}} \bibnamefont{and}
	\bibinfo{author}{\bibfnamefont{A.~V.} \bibnamefont{Sidorov}}, in
	\emph{\bibinfo{booktitle}{{'94 QCD and high-energy hadronic interactions.
				Proceedings, Hadronic Session of the 29th Rencontres de Moriond, Moriond
				Particle Physics Meeting, Meribel les Allues, France, March 19-26, 1994}}}
	(\bibinfo{year}{1994}), pp. \bibinfo{pages}{189--198},
	\eprint{hep-ph/9405254},
	\urlprefix\url{https://inspirehep.net/record/37980/files/arXiv:hep-ph_9405254.pdf}.
	
	\bibitem[{\citenamefont{Kim et~al.}(1998)}]{Kim:1998kia}
	\bibinfo{author}{\bibfnamefont{J.~H.} \bibnamefont{Kim}} \bibnamefont{et~al.},
	\bibinfo{journal}{Phys. Rev. Lett.} \textbf{\bibinfo{volume}{81}},
	\bibinfo{pages}{3595} (\bibinfo{year}{1998}), \eprint{hep-ex/9808015}.
	
	\bibitem[{\citenamefont{Bouchard et~al.}(2017)\citenamefont{Bouchard, Chang,
			Kurth, Orginos, and Walker-Loud}}]{Bouchard:2016heu}
	\bibinfo{author}{\bibfnamefont{C.}~\bibnamefont{Bouchard}},
	\bibinfo{author}{\bibfnamefont{C.~C.} \bibnamefont{Chang}},
	\bibinfo{author}{\bibfnamefont{T.}~\bibnamefont{Kurth}},
	\bibinfo{author}{\bibfnamefont{K.}~\bibnamefont{Orginos}}, \bibnamefont{and}
	\bibinfo{author}{\bibfnamefont{A.}~\bibnamefont{Walker-Loud}},
	\bibinfo{journal}{Phys. Rev.} \textbf{\bibinfo{volume}{D96}},
	\bibinfo{pages}{014504} (\bibinfo{year}{2017}), \eprint{1612.06963}.
	
\bibitem{Liu:2016cba}
L.~Liu {\it et al.},
Phys.\ Rev.\ D {\bf 96} (2017) no.5,  054516.


\bibitem{Luscher:2017cjh}
M.~L\"uscher,
EPJ Web Conf.\  {\bf 175} (2018) 01002.
	
	
\end{thebibliography}
\end{document}